\documentclass[conference]{IEEEtran}
\usepackage{cite}
\usepackage{amsmath,amssymb,amsfonts}
\usepackage{algorithmic}
\usepackage{graphicx}
\usepackage{textcomp}
\usepackage{xcolor}
\usepackage{booktabs}
\usepackage{tablefootnote}
\usepackage{siunitx}
\usepackage[nolist]{acronym}

\begin{acronym}

\acro{3DES}{Triple-DES}

\acro{AES}{Advanced Encryption Standard}
\acro{AI}{Artificial Intelligence}
\acro{ALU}{Arithmetic Logic Unit}
\acro{API}{Application Programming Interface}
\acro{ARX}{Addition Rotation XOR}
\acro{ASIC}{Application Specific Integrated Circuit}
\acro{ASIP}{Application Specific Instruction-Set Processor}
\acro{AS}{Active Serial}
\acro{AWS}{Amazon Web Services}

\acro{BDD}{Reduced Ordered Binary Decision Diagram}
\acro{BFS}{Breadth-First Search}
\acro{BFSM}{Boosted Finite State Machine}
\acro{BGA}{Ball Grid Array}
\acro{BGL}{Boost Graph Library}
\acro{BNF}{Backus-Naur Form}
\acro{BRAM}{block-\acs{RAM}}

\acro{CBC}{Cipher Block Chaining}
\acro{CFB}{Cipher Feedback Mode}
\acro{CFG}{Control Flow Graph}
\acro{CLB}{Configurable Logic Block}
\acro{CLI}{Command Line Interface}
\acro{CMOS}{Complementary Metal-Oxide-Semiconductor}
\acro{COFF}{Common Object File Format}
\acro{CPA}{Correlation Power Analysis}
\acro{CPU}{Central Processing Unit}
\acro{CRC}{Cyclic Redundancy Check}
\acro{CSP}{Cloud Service Provider}
\acro{CTR}{Counter}

\acro{DC}{Direct Current}
\acro{DCP}{Design Checkpoint}
\acro{DDR}{Double Data Rate}
\acro{DES}{Data Encryption Standard}
\acro{DFA}{Differential Frequency Analysis}
\acro{DFT}{Discrete Fourier Transform}
\acro{DFX}{Dynamic Function eXchange}
\acro{DLL}{Dynamic Link Library}
\acro{DMA}{Direct Memory Access}
\acro{DNF}{Disjunctive Normal Form}
\acro{DOS}[DoS]{Denial of Service}
\acro{DPA}{Differential Power Analysis}
\acro{DSO}{Digital Storage Oscilloscope}
\acro{DSP}{digital signal processing}
\acro{DUT}{Device Under Test}
\acro{DRC}{Design Rule Check}

\acro{ECB}{Electronic Code Book}
\acro{ECC}{Elliptic Curve Cryptography}
\acro{ECC}{Error Correction Code}
\acro{EDA}{Electronic Design Automation}
\acro{EEPROM}{Electrically Erasable Programmable Read-only Memory}
\acro{EMA}{Electromagnetic Emanation}
\acro{EM}{electro-magnetic}
\acro{EOP}{Electro Optical Probing}
\acro{EOFM}{Electro Optical Frequency Mapping}

\acro{FA}{Failure Analysis}
\acro{FFT}{Fast Fourier Transformation}
\acro{FF}{Flip Flop}
\acro{FI}{Fault Injection}
\acro{FIB}{Focused Ion Beam}
\acro{FIFO}{First In First Out}
\acro{FinFET}{Fin Field Effect Transistor}
\acro{FIR}{Finite Impulse Response}
\acro{FPGA}{Field Programmable Gate Array}
\acro{FSM}{Finite State Machine}

\acro{GPIO}{General Purpose Input/Output}
\acro{GUI}{Graphical User Interface}

\acro{HDL}{Hardware Description Language}
\acro{HD}{Hamming Distance}
\acro{HDK}{Hardware Development Kit}
\acro{HF}{High Frequency}
\acro{HIL}{High Intensity Light}
\acro{HMAC}{Hash-Based Message Authentication Code}
\acro{HRE}{Hardware \acl{RE}}
\acro{HSM}{Hardware Security Module}
\acro{HW}{Hamming Weight}

\acro{I2C}[I\textsuperscript{2}C]{Inter-Integrated Circuit}
\acro{I2S}[I\textsuperscript{2}S]{Inter-IC Sound}
\acro{IC}{Integrated Circuit}
\acro{ICAP}{Internal Configuration Access Port}
\acro{IIR}{Infinite Impulse Response}
\acro{IO}[I/O]{Input/Output}
\acroplural{IO}[I/Os]{Inputs/Outputs}
\acro{IOB}{Input Output Block}
\acro{IOT}[IoT]{Internet of Things}
\acro{IP}{Intellectual Property}
\acro{IPSW}{iPod Software}
\acro{ISA}{Instruction Set Architecture}
\acro{IV}{Initialization Vector}

\acro{JTAG}{Joint Test Action Group}

\acro{KAT}{Known Answer Test}

\acro{LED}{Light-Emitting Diode}
\acro{LFSR}{Linear Feedback Shift Register}
\acro{LRA}{Linear Resonant Actuator}
\acro{LSB}{Least Significant Bit}
\acro{LUT}{Look-Up Table}
\acro{LVP}{Laser Voltage Probing}

\acro{MAC}{Media Access Control}
\acro{MAC2}[MAC]{Multiply-Accumulate}
\acro{MIPS}{Microprocessor without Interlocked Pipeline Stages}
\acro{ML}{Machine Learning}
\acro{MMIO}{Memory Mapped \acl{IO}}
\acro{MUX}{Multiplexer}
\acroplural{MUX}[muxes]{Multiplexers}
\acro{MSB}{Most Significant Bit}

\acro{NASA}{National Aeronautics and Space Administration}
\acro{NDA}{Non-Disclosure Agreement}
\acro{NMI}{Normalized Mutual Information}
\acro{NSA}{National Security Agency}
\acro{NVM}{Non-Volatile Memory}

\acro{OFB}{Output Feedback Mode}
\acro{OISC}{One Instruction Set Computer}
\acro{ORAM}{Oblivious Random Access Memory}
\acro{OS}{Operating System}
\acro{OSI}{Open Systems Interconnection}

\acro{PAR}{Place-and-Route}
\acro{pBlock}{Dynamic Physical Block}
\acro{PCB}{Printed Circuit Board}
\acro{PC}{Personal Computer}
\acro{PCIe}{Peripheral Component Interconnect Express}
\acro{PDK}{Process Design Kit}
\acro{PEM}{Photon Emission Microscopy}
\acro{PI}{Proportional-Integral}
\acro{PID}{Proportional-Integral-Derivative}
\acro{PL}{Programmable Logic}
\acro{PLB}{Programmable Logic Block}
\acro{PS}{Passive Serial}
\acro{PUF}{Physical Unclonable Function}
\acro{PQC}{Post-Quantum Cryptography}

\acro{RAM}{Random Access Memory}
\acro{RE}{Reverse Engineering}
\acro{RISC}{Reduced Instruction Set Computer}
\acro{RNG}{Random Number Generator}
\acro{ROM}{Read-Only Memory}
\acro{ROP}{Return-oriented Programming}
\acro{RTL}{register transfer level}
\acro{RUB}{Random Unique Block}
\acro{RSA}{Rivest–Shamir–Adleman}

\acro{SAT}{Satisfiability Modulo Theories}
\acro{SCA}{Side-Channel Analysis}
\acro{SEM}{Scanning Electron Microscope}
\acro{SEM}{Soft Error Mitigation}
\acro{SEU}{Single Event Upsets}
\acro{SHA}{Secure Hash Algorithm}
\acro{SHA256}{Secure Hash Algorithm-256}
\acro{SLR}{Super Logic Region}
\acro{SMT}{Satisfiability Modulo Theories}
\acro{SNR}{Signal-to-Noise Ratio}
\acro{SOC}[SoC]{System-on-Chip}
\acro{SPA}{Simple Power Analysis}
\acro{SPI}{Serial Peripheral Interface}
\acro{SRAM}{Static Random Access Memory}
\acro{SRE}{Software \acl{RE}}
\acro{SSI}{Stacked Silicon Interconnect}
\acro{STG}{State Transition Graph}

\acro{UART}{Universal Asynchronous Receiver Transmitter}
\acro{UHF}{Ultra-High Frequency}
\acro{USB}{Universal Serial Bus}

\acro{VHDL}{Very High Speed Integrated Circuit Hardware Description Language}

\acro{WISC}{Writeable Instruction Set Computer}

\acro{XDL}{Xilinx Description Language}
\acro{XST}{Xilinx Synthesis Technology}
\acro{XTS}{XEX-based Tweaked-codebook with ciphertext Stealing}

\acro{FAR}{Frame Address Register}
\acro{FDRI}{Frame Data Register Input}
\acro{FDRO}{Frame Data Register Output}
\acro{CTL0}{Control Register 0}
\acro{COR0}{Configuration Register 0}
\acro{CTL1}{Control Register 1}
\acro{COR1}{Configuration Register 1}

\acro{PR}{Partial Reconfiguration}
\acro{ICAP}{Internal Configuration Access Port}
\acro{PL}{Programmable Logic}
\acro{PoC}{Proof-of-Concept}

\end{acronym}

\usepackage{hyperref}
\hypersetup{%
  colorlinks=true,%
  linkcolor={black},
  citecolor={black},
  urlcolor={black},
  bookmarksnumbered=true,%
  bookmarksopen=true}

\def\BibTeX{{\rm B\kern-.05em{\sc i\kern-.025em b}\kern-.08em
    T\kern-.1667em\lower.7ex\hbox{E}\kern-.125emX}}
\begin{document}

\title{The Achilles' Heel of Partial Reconfiguration: Optical Side-Channel Leakage on the 7-Series ICAP}

\author{
\IEEEauthorblockN{Antonio Saavedra, Jan Caspar Marx, Lars Renkes, Jean-Pierre Seifert}
\IEEEauthorblockA{\textit{Security in Telecommunications, Technische Universität Berlin, Berlin, Germany} \\
\{a.saavedra, j.marx, lars.renkes, jean-pierre.seifert\}@tu-berlin.de}
}

\maketitle

\begin{abstract}
Major FPGA manufacturers have incorporated bitstream encryption to protect sensitive configuration data. However, for the most widely used FPGA families, multiple attacks against unpatchable protection schemes hard-wired into the devices can bypass or fully break them, making patchable schemes desirable.

In this work, we present a proof-of-concept implementation of an AMD-proposed asymmetric key encryption scheme for bitstream protection for 7-Series FPGAs, using partial reconfiguration from the Programmable Logic. We analyze the security implications and hardware overhead of this implementation. We then propose and demonstrate an optical side-channel attack that is able to recover plain-text configuration data during the dynamic reconfiguration process. This attack leverages Photon Emission Microscopy and Electro-Optical Probing to first locate and then contactlessly extract the plain-text data from the ICAP interface, which internally connects the Programmable Logic with the configuration logic.

We located the ICAP buses in an AMD XC7A200T device and show that the data on it can be extracted with Electro-Optical Probing. We claim that even advanced encryption schemes utilizing Partial Reconfiguration and custom cryptographic engines are vulnerable to optical attacks, as reconfiguration is only possible via hard-wired, vulnerable configuration interfaces.
\end{abstract}



\section{Introduction}
Bitstream protection is a cornerstone of FPGA security as it protects sensitive information and intellectual property~\cite{DBLP:journals/trets/ProulxCFM23}.
Therefore, all major FPGA manufacturers have incorporated hard-wired cryptographic engines into their products to support configuration data protection schemes.

At the same time, protection mechanisms have been the target of multiple attacks.
In particular, the encryption mechanisms in AMD 7-Series FPGAs have been successfully broken in previous works~\cite{CCS:MBKP11,COSADE:MorSch16}.
These schemes consists of decryption engines (such as \ac{AES} in the case of 7-Series FPGAs) that are embedded in the configuration logic and therefore cannot be patched or upgraded after manufacturing, as highlighted by Ender et. al.\cite{USENIX:EndMorPaa20}.
In response, multiple patchable protection schemes have been proposed~\cite{DBLP:conf/ccs/UntersteinJHGH19,DBLP:journals/cryptography/OwenHCCSAP18,DBLP:journals/trets/KashyapC16,WP468}.
These works implement bitstream protection mechanisms in \ac{PL} and are, in contrast, patchable.

\begin{figure}[t]
	\centering
	\includegraphics[width=0.9\linewidth]{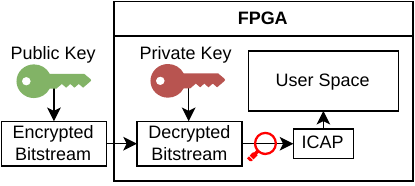}
	\caption{High-level overview of the Asymmetric Bitstream Encryption Scheme.}
	\label{fig:scheme}
\end{figure}

In the same line, AMD's \acl{DFX} User Guide, UG909~\cite{UG909}, proposes an asymmetric key encryption scheme based on their \ac{PR} capabilities.
The main concept of the scheme, shown in \autoref{fig:scheme}, is the use of asymmetric key encryption for the bitstream, with on-chip key generation and bitstream decryption.
A trusted host, e.g. a trusted computer, can obtain the public key from the FPGA, and use it to encrypt a partial bitstream containing the user design.
This encrypted partial bitstream can be sent back to the FPGA, where it is decrypted and is used to perform \ac{PR} through the \ac{ICAP}.
The implementation of this scheme splits the FPGA into a static part, where the hardware required for key generation and decryption is implemented, and a dynamic user space.
The static part also contains the logic to interact with the configuration logic through the \ac{ICAP} port and the communication interface with a host.
AMD claims that one of the main advantages of this scheme is that, even if the device is stolen and remains powered on, the private key is extremely difficult to recover.
As the key is stored in \ac{PL} memory, it does not have a fixed physical location, in contrast to the eFUSEs or BBRAM used to store the embedded AES engine's key.

In this work, we developed and tested a \ac{PoC} implementation of this asymmetric key encryption scheme to later demonstrate that an optical side-channel attack is capable of recovering plain-text configuration data.
Our attack uses \ac{PEM} to locate the target \ac{ICAP} interface and map each of the interface pins. Once the interface is located and mapped, we perform \ac{EOP} to recover the waveform of the data that traverse it.
We confirmed that the probed waveforms are plain-text configuration data. 
The location vulnerable to optical probing is highlighted in \autoref{fig:scheme} with a magnifying glass.

The configuration logic has been a target of optical side-channels attacks before~\cite{CCS:TLSB17, DBLP:journals/jhss/KrachenfelsLSDF20}. 
However, to the best of our knowledge, this is the first attack that exploits optical side-channels on the \ac{ICAP} interface.
With our attack, we demonstrate that even patchable bitstream protection schemes are vulnerable to optical side-channels attacks, as the configuration logic requires plain-text data from the \ac{PL} when performing \ac{PR} with custom decryption cores.
Even with patchable schemes, the hard-wired configuration interfaces remain an Achilles' heel, compromising an otherwise robust implementation.

Our contributions can be summarized as follows:
\begin{itemize}
\item We implemented an AMD-proposed secure scheme~\cite{UG909} based on asymmetric key encryption and partial reconfiguration.
\item We attacked this implementation with an optical side-channel exploit, detecting and mapping the vulnerable structures on the configuration logic.
\item We analyzed the security implications of this attack surface, proposed practical countermeasures, and outlined directions for future work.
\end{itemize}

We disclosed our findings to AMD in May 2026. AMD acknowledged our work, but they do not consider physical backside attacks within their threat model. The complete source code, including the RTL design and host-side scripts, is publicly available at \url{https://github.com/JnCrMx/icap-probing-poc}.


\section{Background}
\label{sec::background}

\begin{figure*}[ht]
	\centering
	\includegraphics[width=\linewidth]{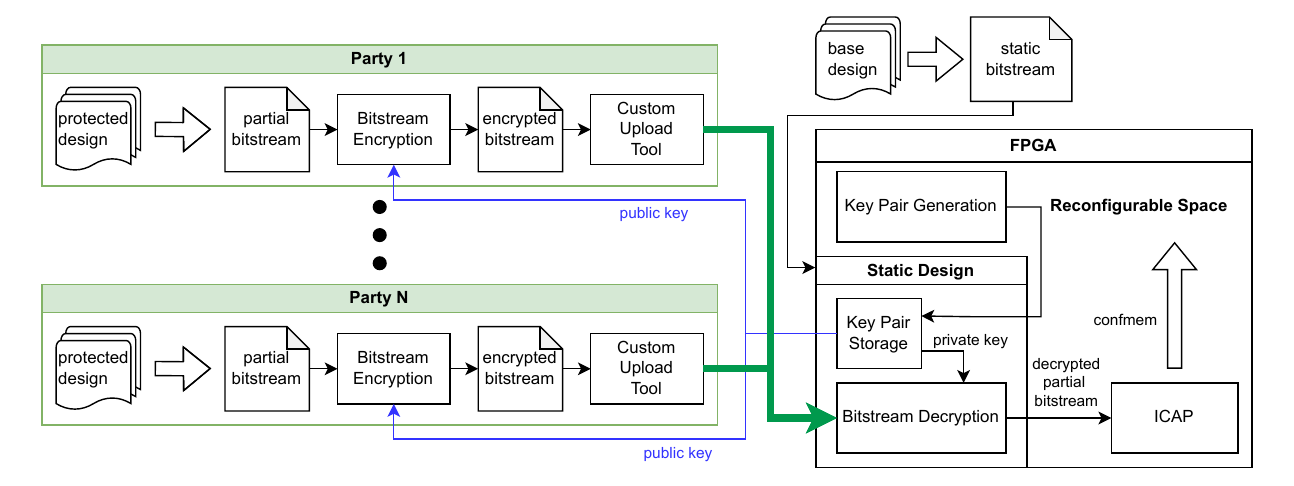}
	\caption{Block Diagram of the Asymmetric Bitstream Encryption Scheme and Multi-Party Configuration Flow.}
	\label{fig:high-level}
\end{figure*} 

\subsection{Partial Reconfiguration}
Modern FPGA architectures support dynamic \acf{PR}.
This capability allows devices to split their \ac{PL} resources into different regions, which can be reprogrammed with separated bitstreams during runtime.
For AMD FPGAs, the device fabric is partitioned into static and reconfigurable regions, called Partial Blocks (P-Blocks).
A P-Block encapsulates a specific subset of \acp{CLB}, DSP slices, and BlockRAM tiles, along with the necessary routing resources to isolate it from static logic.
During runtime, the FPGA can load a partial bitstream for a specific P-Block, without disrupting the functionality of the other regions.
Just as for regular bitstreams, the process of loading a partial bitstream is controlled by the configuration logic, which can be accessed by multiple programming interfaces.

For the proposed asymmetric key bitstream protection scheme, as the reconfiguration process is controlled from the \ac{PL}, the \ac{ICAP} is needed.
The \ac{ICAP} is a hardware interface that connects the configuration logic with the \ac{PL}.
It enables read and write access to the configuration memory internally from the \ac{PL}, without requiring access to external configuration interfaces.
In 7-Series devices, the \ac{ICAP} interface is up to 32-bit wide and can operate with maximum clock frequencies from 100 MHz to 200 MHz depending on the device family and speed grade. 


\subsection{Optical Side-Channels}
\label{subsection::background::optics}
This work uses \ac{PEM} and \ac{EOP}, two contactless optical techniques widely applied in integrated circuit failure analysis, providing insight into the internals of an IC through the silicon backside.
Both have been shown to enable optical side-channel attacks in hardware security research~\cite{DBLP:conf/host/CarmonSW17, TCHES:KRSSHA24}.
In addition, both techniques are combined with infrared imaging to capture structural backside images of integrated circuits.

\subsubsection{Photon Emission Microscopy}
Photon emission in semiconductors originates from electron-hole recombination~\cite{pem}.
In CMOS devices, photon emission occurs during switching events when current flows, making the emission intensity proportional to switching activity and current density~\cite{pem_cmos}.
The emitted photons are recorded with an InGaAs camera, which integrates the measurements over a defined period.
The resulting image is an activity heatmap that can be overlayed onto the structural pattern image.

\subsubsection{Electro Optical Probing}
\ac{EOP} enables contactless backside probing of integrated circuits.
Infrared light is focused onto the active device area, where voltage-dependent free-carrier effects modify the optical properties of the material and, thus, the reflected light intensity~\cite{invest_of_lvp}.
As these intensity changes are very small, many synchronized measurements must be averaged to achieve sufficient \ac{SNR}~\cite{lvp}.
Besides incoherent light sources, coherent laser sources can also be used, known as \ac{LVP}.
The result of an \ac{EOP} measurement is a waveform that represents the voltage over the measurement time at the specific probing location.

\subsection{Related Work}

As mentioned previously, bitstream protection is one of the most researched topics in FPGA security.
Access to plain-text configuration data could lead an attacker to IP theft, reverse engineering, or insertion of malicious hardware trojans~\cite{DBLP:journals/trets/ProulxCFM23}. 
The first attacks on bitstream protection schemes for 7-Series devices, which date back to 2011, focused on power side-channels to extract AES keys~\cite{CCS:MBKP11}. Since then, newer side-channel analysis techniques have further improved this attack vector~\cite{COSADE:MorSch16,DBLP:journals/tc/RamezanpourAD20}.

More recently, interest has shifted to design weaknesses. The Starbleed attack~\cite{USENIX:EndMorPaa20} fully breaks  encryption on 7-Series devices, by storing the plain-text configuration words, after decryption, in an internal register from the configuration logic.
As this register is persistent memory, storing a decrypted configuration word into it allows it to be read out externally after a reboot.
JustSTART~\cite{TCHES:EHFMP24}, a further attack based on design weaknesses, shows how the newer Bitstream authentication on Ultrascale(+) devices can be bypassed.
Both of these attacks demonstrate structural weaknesses of the implementation of the configuration logic that, as such, are unpatchable. This highlights the need for patchable security mechanisms.

The \ac{ICAP} has also been shown to have structural weaknesses in security. STRAT~\cite{DBLP:journals/trets/AlbartusEMFPT24} is a framework for malicious partial bitstream injection into configuration memory by exploiting \ac{ICAP} vulnerabilities.
The vulnerability of reconfiguring through the ICAP is rooted in the configuration logic directly programming the partial bitstream, without any verification of which configuration frames are overwritten.
This enables, for example, the injection of trojans into other P-Blocks or the malicious modification of the static design.

Our work extends the attack vectors for the \ac{ICAP}, as we demonstrate that, besides the implementation weaknesses of the \ac{PR} process, the \ac{ICAP} also presents side-channel leakage through optical channels, which enables non-invasive probing.

Besides power side-channels and implementation attacks, contactless probing techniques have been proven capable of reading out signals from FPGAs.
An attack using optical probing~\cite{CHES:LTBS16} could successfully read signals from the \ac{PL} on 60-nm FPGA devices.
Keys stored on an Ultrascale's BBRAM have been read with laser voltage stimulation using a low-cost setup~\cite{DBLP:journals/jhss/KrachenfelsLSDF20}.
The bitstream protection on the configuration logic itself has also been the target of optical-side channel attacks.
Using \ac{EOFM} and custom bitstreams, a profiling attack~\cite{CCS:TLSB17} could identify internal connection buses that carry the plain-text configuration data after decryption.
The Leakback CRC attack~\cite{LeakbackCRC} demonstrated that using \ac{EOP}, the Readback CRC capabilities can be exploited to read the plain-text configuration data from internal buses in encrypted designs. 

Our work expands on previous attacks, as, to the best of our knowledge, no optical probing attack has been targeted at the \ac{ICAP} interface, highlighting the impact of the optical side-channel for the privacy of the \ac{PR} process.

Other related bitstream protection schemes have been proposed, particularly for Zynq \acp{SOC}. A framework\cite{DBLP:journals/trets/KashyapC16} was developed to protect partial bitstreams for \ac{PR}, which uses randomly generated keys on the device to encrypt partial bitstreams before storing them in non-volatile memory. 
For \ac{PR}, the partial bitstream is then decrypted and programmed through the \ac{ICAP}. A self-authenticating framework~\cite{DBLP:journals/cryptography/OwenHCCSAP18} was also developed to secure a two-stage boot process on a Zynq device. 
First, a non-protected bitstream is loaded into the \ac{PL}, authenticates itself, and uses a \ac{PUF} to generate a decryption key. 
With this key, a second protected bitstream can be decrypted in the PL and programmed via \ac{PR}. Another patchable scheme~\cite{DBLP:conf/ccs/UntersteinJHGH19} uses a \ac{PUF} to embed a secret key into the \ac{PL}, which is then used to decrypt protected bitstreams. The decrypted bitstream is then used to perform \ac{PR} through the \ac{ICAP}. 
All of these works transmit sensitive plain-text configuration data through the \ac{ICAP}, making them vulnerable to optical side-channel leakage.


\section{PoC implementation}
\label{sec::implementation}
\begin{figure*}[ht]
	\centering
	\includegraphics[width=\linewidth]{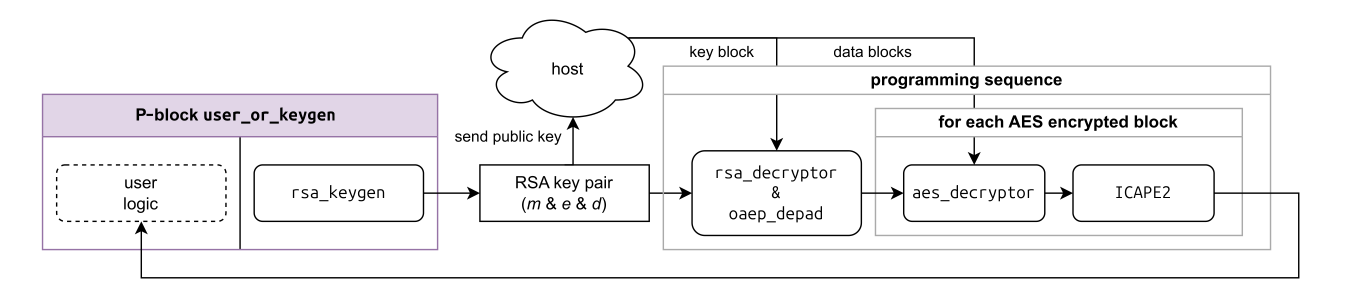}
	\caption{Data Flow Diagram of the Hybrid Encryption Scheme.}
	\label{fig:dataflow}
\end{figure*}

To showcase our attack, we have implemented the asymmetric bitstream encryption scheme proposed by AMD ~in UG909~\cite{UG909}, which is described as a ``very secure method for protecting the FPGA configuration file''.

\subsection{Design Goals}

This encryption scheme allows an arbitrary number of parties to generate and encrypt bitstreams for the same FPGA, as soon as they receive the FPGA's public key over a secure channel.
This scheme is shown in \autoref{fig:high-level}.
In this case, the main advantage of an asymmetric encryption system in place of a symmetric one is that the parties do not have to trust each other.
There are no secrets shared between the different parties, which guaranties that even if any party is fully compromised (or becomes an adversary), all bitstreams encrypted by the other parties (both in the past and the future) remain fully secure.
Hence, this system is designed to protect against leaks of encrypted partial bitstreams, hostile other parties, and physical theft of the FPGA.

\subsection{Overview}

Our implementation uses a static partition for the control, communication, and decryption logic as well as a single dynamic partition (\texttt{user\_or\_keygen}) for key generation and user logic.

On startup, the reconfigurable partition \texttt{user\_or\_keygen} initially contains the RSA key pair generation logic, which immediately starts to generate an RSA key pair.
Once the key pair has been generated, the system is ready for operation, and, as not needed anymore, key generation logic can be replaced with user designs.

Any host can now request the public key by issuing a command over a serial connection.
Furthermore, any host can initiate the programming sequence by sending the appropriate command.
During the programming sequence, the host first sends an RSA encrypted key block, which contains a randomly chosen AES key.
The FPGA will then use this AES key to decrypt the actual bitstream blocks, which are sent one by one by the host and immediately forwarded to the ICAP after decryption. The data flow of our implementation is shown in \autoref{fig:dataflow}. 
A more detailed explanation about this hybrid encryption scheme follows in subsection D.

\subsection{Key Generation}
RSA key generation takes place fully on the FPGA itself with random bit values provided by a combination of multiple ring oscillators based on the design proposed by Sunar et al.~\cite{SunarRng}.
The key generation logic is located in the reconfigurable partition of the design and will be overwritten by the user-supplied design after the first partial bitstream programming took place.
This is done because the key pair only needs to be generated once after the programming of the initial full bitstream and placing it inside the reconfigurable partition reduces the area utilization of the static design, leaving more area available for user designs.

The generated key pair is stored in \acp{CLB} registers, and only the public key is made available to external devices via the host interface.
This ensures that the private key cannot be easily retrieved by any other device, even if they have physical access to the FPGA.
Should the FPGA lose power, the private key is immediately lost by design and cannot be recovered.
In this case, the FPGA would generate an entirely new random RSA key pair upon reprogramming, and the public key needs to be retrieved by all involved parties again.

\subsection{Bitstream Encryption and Decryption}
The partial bitstreams are encrypted using a hybrid encryption scheme based on \acs{RSA} and \ac{AES}.
Although using a full asymmetric encryption scheme is theoretically possible, for our \ac{PoC} implementation, we opted for the hybrid scheme to speed up the programming process.

To encrypt a bitstream, a host generates a random AES-256 key and \ac{AES}-\acs{CBC} \ac{IV}, and uses them for bitstream encryption.
It stores the \ac{AES} key and \ac{IV} together in a \emph{key block}, which it then encrypted with the FPGA's public key using RSA-OAEP.

To program an encrypted bitstream, the host first sends the RSA encrypted key block, which the FPGA then decrypts and depads.
The FPGA checks that the decryption was successfully (which can be determined based on the correctness of the padding), initializes its AES decryption module with the decrypted AES key and IV, and reports success to the host.
Now, the host can send the encrypted bitstream block-by-block.
The FPGA decrypts each block and sends it directly to the ICAP as 32-bit words.

\subsection{Host Interface}
In our \ac{PoC} implementation, the target FPGA and the host communicate via a USB-UART bridge with a baud rate of 921600.
A custom simple communication protocol was implemented, where the host sends a single command byte to the FPGA beginning the execution of one of the two available commands:
\begin{itemize}
    \item 'get public key': This command requests the public key from the device. The device responds by the size of the public key followed by the public key data.
    \item 'program': This command begins the programming sequence: The host immediately sends the number of AES encrypted bitstream blocks followed by the RSA encrypted key block, after which the device decrypts and confirms the key block. The encrypted bitstream blocks are then sent one at a time. After each block has been decrypted and forwarded to the ICAP, the FPGA sends a single byte as confirmation, instructing the host to send the next block.
\end{itemize}

\subsection{Timing and Utilization}
Our design supports two different RSA key lengths of \SI{768}{\bit} and \SI{1024}{\bit}. The timing and utilization depend on the chosen key length. \autoref{tab:area} shows the LUTs and FFs used for the static design and key generation logic, as well as the rounded percentage of the total available on the device. As after initialization the key generation logic in the P-Block can be overwritten by the user design, for both key lengths we configured this P-block to spread over 5 out of 10 clock regions (X1Y0 to X1Y4), which contains approximately 51\%, 88\% and 44\% of the available Slices, BRAM and DSPs, respectively. Further resource optimization is possible, as the static design needs at most 35\% of the available Slices. 

\begin{table}[t]
    \centering
    \caption{Resource utilization of the PoC implementation.}
    \small
    \label{tab:area}
    \begin{tabular}{lcc}
        \toprule
        \textbf{RSA Key Length} & \textbf{768 bit} & \textbf{1024 bit} \\
        \midrule
        Static Design Slice LUTs    & 34,908 (25\%)& 45,872 (34\%)\\
        Static Design Slice FFs     & 33,361 (12\%)& 42,954 (16\%)\\
        Key Generation Slice LUTs   & 30,691 (23\%)& 41,547 (31\%)\\
        Key Generation Slice FFs    & 20,434 (8\%)& 27,114 (10\%)\\
        \bottomrule
    \end{tabular}
\end{table}

Our design contains three different clock domains: $f_{\text{ctrl}}$ for control, host communication, ICAP interface and AES decryption, $f_{\text{RSA,keygen}}$ for RSA key generation and $f_{\text{RSA,decryptor}}$ for RSA decryption.
We set the three clock frequencies to be integer multiples of each other to avoid clock domain crossings.
\autoref{tab:timing} shows the frequency for each domain for the two different key lengths.
Implementing proper clock domain crossing could increase the AES decryption and control logic speed by around \SI{25}{\percent}, however, we have decided that for this \ac{PoC} the speedup does not justify the added complexity.

In addition, \autoref{tab:timing} displays the latency of the system startup and RSA decryption.
$\bar{t}_{startup}$ describes the average duration of the startup sequence after the static bitstream has been programmed, which consists of RSA key pair generation and RSA decryption logic initialization (pre-computing values that are independent of input data).
As the startup time is not constant and depends on how quickly the key generation logic can find two random prime numbers, we averaged 200 measurements for each key length on the target FPGA.

$\bar{t}_{RSA,decryption}$ describes the average duration of the RSA decryption operation, which takes place at the beginning of each programming sequence to decrypt the RSA encrypted key block containing the AES key.
As the decryption time in our implementation only depends on the private key, we simulated this process for 32 different pairs of RSA keys and averaged the required time.

\begin{table}[h]
    \centering
    \caption{Timing characteristics of the implemented scheme.}
    \small
    \begin{tabular}{lcc} 
        \toprule
        \textbf{RSA Key Length} & \textbf{768 bit} & \textbf{1024 bit} \\
        \midrule
        $f_{\text{ctrl}}$ & \SI{60}{\mega\hertz} & \SI{40}{\mega\hertz} \\
        $f_{\text{RSA,keygen}}$ & \SI{30}{\mega\hertz} & \SI{20}{\mega\hertz} \\
        $f_{\text{RSA,decryptor}}$ & \SI{15}{\mega\hertz} & \SI{10}{\mega\hertz} \\
        $\bar{t}_{\text{startup}}$ & \SI{7.82}{\second} & \SI{24.19}{\second} \\
        $\bar{t}_{\text{RSA,decryption}}$ & \SI{59.22}{\milli\second} & \SI{157.85}{\milli\second} \\
        \bottomrule
    \end{tabular}
    \label{tab:timing}
\end{table}

\section{Attack}

\begin{figure*}[ht]
	\centering
	\includegraphics[width=\textwidth]{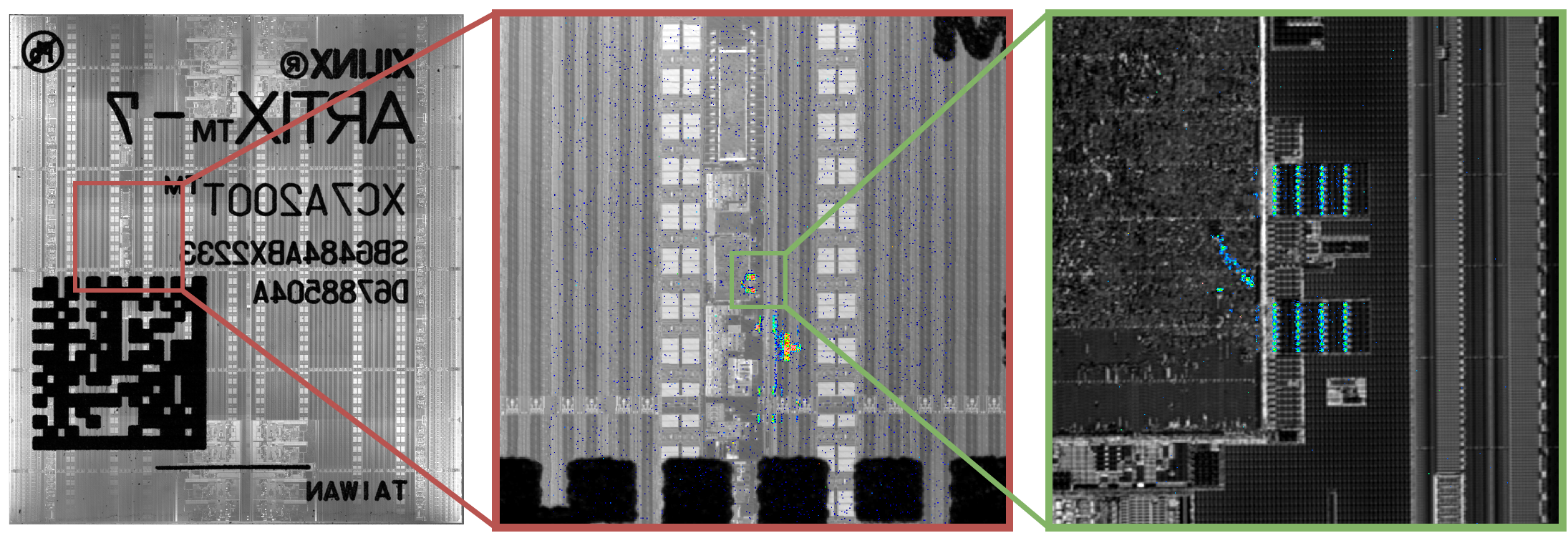}
	\caption{Backside Infrared Images of the XC7A200T. From Left to Right: Entire Device;  5x Magnification; and 50x Magnification with superimposed PEM.}
	\label{fig:device}
\end{figure*}
\label{sec::attack}

We demonstrate our attack on the \ac{PoC} implementation using an RSA key length of \SI{768}{\bit}.
The attack can be separated into two steps. First, the exact location of the ICAP buses must be found, and each pin must be properly mapped. Then, the data can be probed, for which a malicious host is required.

\subsection{Device under attack and equipment}
For our experiments, we targeted an XC7A200T Artix-7 FPGA on the Digilent Nexys Video Development Board. This device is packaged in an SBG484 package, which is a lidless flip-chip with \ac{BGA} connections. As it is a flip-chip, the backside bulk silicon of the device is directly underneath a glued heat sink that the Nexys Video board incorporates. Removing this heat sink directly exposes the bulk silicon. This device, as well as other lidless packages, has a QR-Code and serial number laser-engraved to the silicon surface. However, since they do not cover the \ac{ICAP} interface, for our purposes, the device does not require further preparation after removing the heat sink.

The optical equipment we use for our experiments is a Hamamatsu PHEMOS-1000 emission microscope. It includes a C8250-27 InGaAs camara, which we used to perform \ac{PEM} measurements. Infrared pattern images and \ac{EOP} measurements were obtained with a \SI{1300}{\nano\meter} \ac{HIL} that is fitted to the system. 

\subsection{Finding ICAP location}

\begin{figure*}[ht]
	\centering
	\includegraphics[width=\textwidth]{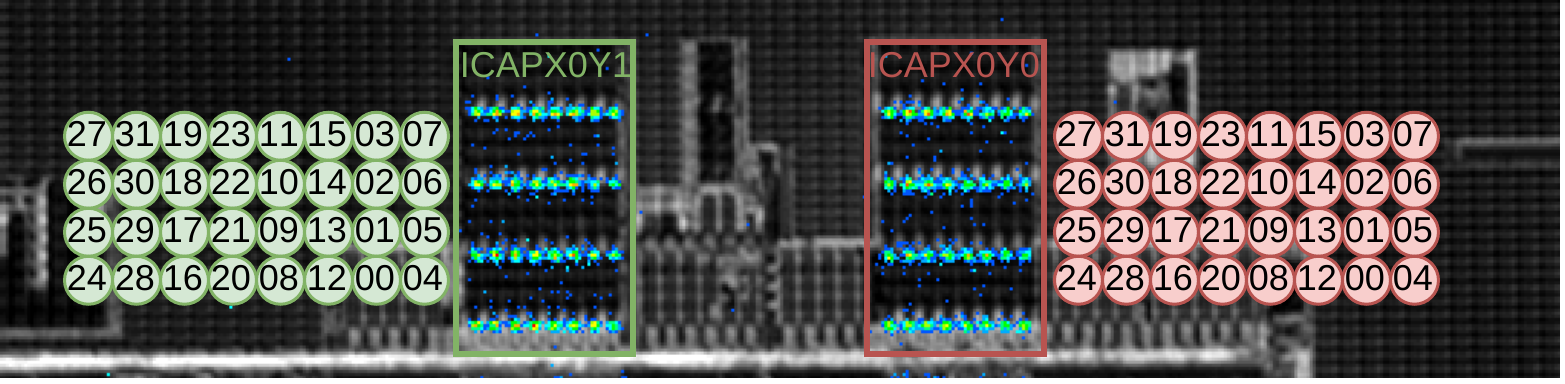}
	\caption{PEM overlaid on 50x Infrared Image of the ICAPX0Y1 and ICAPX0Y0 Interfaces of the XC7A200T device with the mapping of the bit order.}
	\label{fig:bitmapping}
\end{figure*}

In order to identify the \ac{ICAP} interface, we used \ac{PEM}. For this purpose, we developed an FPGA design that injects switching activity into the \ac{ICAP} interface. As explained in \autoref{subsection::background::optics}, the amount of photon emission is proportional to the transistor switching activity. Our target device has two \ac{ICAP} interfaces that can be clocked at up to 100 MHz. Our design writes to both \ac{ICAP} interfaces, using the Vivado ICAPE2 primitive, alternating 32-bit words of all 1s and 0s. This allows us to inject the highest possible switching activity.

\autoref{fig:device} shows the entire device and the photon emission hotspots in the configuration logic at 5x and 50x magnification. We found that the two interfaces are next to each other at the edge of the configuration logic. The 32 bits of the \ac{ICAP} input interface, with write access from the \ac{PL} to the configuration logic, are ordered in an $8\times4$ array. In \autoref{fig:device}, the photon emission is overlayed onto the black-and-white infrared pattern image, and the 32 hotspots visible for each interface correspond to their 32 bit.

In order to map this $8\times4$ array, we modified our design to only switch one bit at a time, allowing us to detect the order of the layout. \autoref{fig:bitmapping} shows the mapping of all 32 input pins to each \ac{ICAP}. Our experiments show that the mappings are the same for both \acp{ICAP}, labeled in Vivado as ICAPX0Y0 and ICAPX0Y1 for our device. 

\subsection{Malicious Host}
As in the implemented PoC protection scheme, multiple parties can program an encrypted configuration bitstream to the device under attack, we set off the scenario that one host has been compromised.
The attack is then carried out from this malicious host.
\ac{EOP} requires multiple integrations of the same data, which in turn requires a periodic trigger signal aligned with that data.
Therefore, the main task of a malicious host is to generate periodicity in the target data and a trigger signal aligned to it. 
We implemented this malicious host on an external computer that communicates with the device-under-attack through the USB-UART bridge on the Nexys Video board.
The malicious host needs only access to an encrypted bitstream and an active connection to the device under attack.
The malicious host can then send the encrypted bitstream fully or partially to the target in a loop.
For our attack, we set the malicious host to repeatedly send the first encrypted configuration data block in a continuous loop.


\subsection{Electro-Optical Probing}

\label{sec::results}
\begin{figure}[ht]
	\centering
	\includegraphics[width=\linewidth]{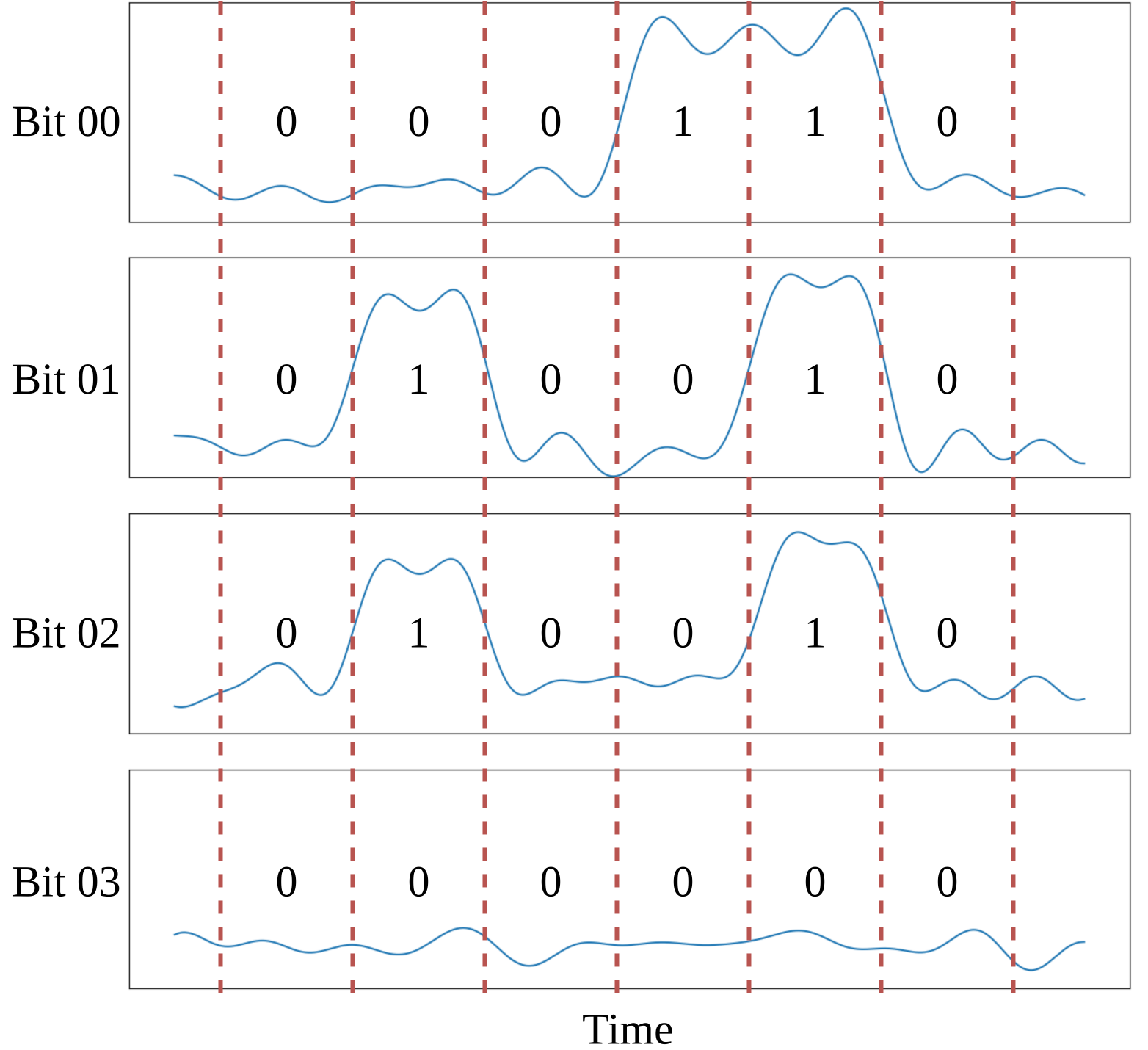}
	\caption{EOP Waveforms of the four LSBs of the first configration block.}
	\label{fig:probed-waveforms}
\end{figure}

Using \ac{EOP}, we probed the signal traversing the ICAPX0Y1.
\autoref{fig:probed-waveforms} shows \ac{EOP} waveforms obtained from the first decrypted configuration block. These waveforms plot the estimated voltage over time. With a simple thresholding algorithm, the values can be binarized to logic 0s or 1s. To help visualize the data, we added red dashed lines to separate different clock cycles. As the malicious host sets the trigger to start measuring shortly before the first data is transmitted through the \ac{ICAP}, the first clock cycle shown in the waveforms is prior to the actual data, which starts on the second clock cycle shown.

For simplicity, \autoref{fig:probed-waveforms} only shows the four least significant bits.
In our test bitstream, as well as in most Vivado-generated partial bitstreams, the first configuration words correspond to xAA995566, x20000000, x30000801, and x00000007.
These configuration words are, respectively, the synchronization word, a NOP, and a write to the CMD register.
Accordingly, the four least significant bits of these words are 0110, 0000, 0001 and 0111 which match the probed data shown in \autoref{fig:probed-waveforms}, when read vertically.


\section{Discussion}
\subsection{Limitations}
Although our attack successfully demonstrates the feasibility of extracting plain-text configuration data via the ICAP interface, there are limitations to this attack vector.

For our \ac{PoC} implementation, programming the complete user bitstream takes around \SI{50}{\second}, due to the low UART data rate and the delays introduced by AES-CBC. 
This results in years of measurement time to recover the complete user bitstream.
However, AES-CBC allows an attacker to only program and probe one block at a time, which only takes around \SI{4}{\hour} of measurement time.
An attacker does not necessarily need the entire bitstream to compromise a system, as sensitive data can reside in specific configuration frames.
For example, initial BRAM values are stored in a specific section of the bitstream. An adversary could selectively target these sections. Thus, while full bitstream recovery is currently impractical, the exposure of confidential data remains a viable threat.

The total measurement time is dominated by the use of block ciphers, as data do not flow continuously through the ICAP. For comparison, LeakbackCRC~\cite{LeakbackCRC} demonstrated that the bitstream of a small RISC-V core can be recovered in \SI{40}{\hour}.

A further limitation is the access to optical probing tools.
\ac{PEM} and \ac{EOP} require specialized equipment that is typically used for fault analysis.
\ac{PEM} is only needed if the location of the \ac{ICAP} is not known.
Alternatively, as all 7-Series devices have similar layouts, it can be located by visual inspection using our results (\autoref{fig:device}) as a reference. We tested this by locating the \ac{ICAP} in an XC7K325T device by visual inspection and confirmed the bitmapping by \ac{PEM}.

Furthermore, the attack requires access to the silicon backside of the die.
For flip-chip packages, this access is relatively straightforward.
From the AMD portfolio, only the lower-end devices, like the smaller Artix 7, are not packed as a flip-chip.
Lidless flip-chips, as our XC7A200T-1SBG484C, feature a laser-engraved QR code on the bulk silicon backside.
These engravings can be removed by mechanical polishing in case they are covering the targeted structures.

\subsection{Extent of the vulnerability}

The optical side-channel on the \ac{ICAP} bus is not limited to our proof-of-concept implementation.
Other protection schemes, which also decrypt the bitstream on the \ac{PL} and rely on the \ac{ICAP} for \ac{PR}, could also be vulnerable. 
Among these schemes, implementations for Zynq \acp{SOC}~\cite{DBLP:conf/ccs/UntersteinJHGH19,DBLP:journals/cryptography/OwenHCCSAP18,DBLP:journals/trets/KashyapC16,WP468} also use the \ac{ICAP} to transmit plain-text configuration data, making them vulnerable to optical probing attacks.

AMD has incorporated more advanced security measures into newer devices. In UltraScale(+) devices, asymmetric key encryption is used instead of \ac{AES}. 
Spartan UltraScale+ introduced a new configuration logic and is expected to incorporate \ac{PQC} capabilities in the future. Although access to the new configuration logic appears to be limited to IPs and Soft-CPUs, the \ac{ICAP} seems to still be present for the \ac{SEM} core. At the time of writing, the first devices have recently launched and only limited documentation and public knowledge are available.

Furthermore, Ultrascale(+) devices have migrated to smaller process nodes, namely \SI{20}{\nano\meter} for Ultrascale and \SI{16}{\nano\meter} for Ultrascale+.
The 7-Series devices targeted in this work utilize a \SI{28}{\nano\meter} process.
This scaling pushes the limits of the spatial resolution achievable with standard optical techniques.
However, Electron-Beam Probing, although being more expensive and requiring invasive preparation, has been shown to be capable of extracting signals from Ultrascale+ devices~\cite{saavedra_exploring_2024}.

\subsection{Countermeasures}

One potential countermeasure is to clock the \ac{ICAP} interface with an irregular clock source or with significant clock jitter.
Since \ac{EOP} takes multiple measurements and integrates them over time, these measurements need to be aligned with a fixed offset to a shared trigger signal.
An irregular clock moves this offset unpredictably.
Similarly, adding random delays in between \ac{ICAP} writes would hinder proper alignment of subsequent \ac{EOP} traces.
The bitstream can also be written out of order, as writing to the \ac{FAR} allows to control to which frame is being written.
This would not eliminate the attack vector entirely, as any write to the \ac{FAR} would also leak through the \ac{ICAP}, but it would significantly increase the complexity of the data reconstruction.

In general, there are possible countermeasures that manufacturers can include in their devices, to protect hard-wired programming interfaces from optical side-channels.
An approach is to use sensors in sensitive locations to detect physical changes due to probing attacks~\cite{lvp_cost} or to protect the device's backside with a reflective coating, completely preventing backside access~\cite{DBLP:journals/jhss/AminiBHSSB18}.
LaserEscape~\cite{DBLP:conf/iccad/MonfaredMCFT24} is an \ac{EOP} countermeasure, which can be incorporated by users into custom designs, that leverages \ac{PR} to relocate the user design when probing is detected. However, it still relies on the ICAP to perform \ac{PR} and can only detect probing attacks targeting the \ac{PL}.

\subsection{Future Work}
In future work, we consider enhancing the security of \ac{PR} through the \ac{ICAP} to partially patch its vulnerabilities.
For example, the addresses of the configuration frames to be transmitted can be checked, filtering out potential malicious bitstreams that write to frames that should be protected.
Some of our proposed countermeasures, such as introducing random delays and modifying the order in which the frames are written, can be incorporated into these protection mechanisms. 

In addition, this attack surface should be studied in the new Spartan Ultrascale+ FPGAs.
Public commercial information on the device states that they incorporate enhanced security mechanisms, including \ac{PQC}. At the time of writing, high-end Spartan Ultrascale+ devices, which include \ac{PQC} capabilities (from SU45P to SU200P), are not yet commercially available. In the same line, the new configuration logic, which has been updated from the 7-Series and previous Ultrascale(+) devices, makes them an interesting target to study.


\section{Conclusions}
\label{sec::conclusions}

In this work, we expose the optical side-channel leakage of the \ac{ICAP} interface in 7-Series FPGAs. For this purpose, we implemented an asymmetric key encryption scheme for bitstream protection, proposed by AMD in UG909, and presented an attack that exploits optical side-channel leakage of the \ac{ICAP} to recover plain-text configuration data.

Our \ac{PoC} implementation of the patchable requires \SI{50}{\second} to fully configure the available user space. The available resources in this user space are 51\% of the total LUTs and FFs, as well as 88\% of the BRAM and 44\%  of the DSP Slices. 

By combining \ac{PEM} to locate and map the internal configuration interface, with \ac{EOP} to perform contactless probing, we successfully recovered plain-text data traversing the \ac{ICAP}, bypassing the encryption scheme entirely.
This finding exposes the internal programming interfaces as a source of optical side-channel leakage, calling for a reevaluation of patchable encryption schemes for FPGAs, as to the best of our knowledge, all published implementations utilize these hard-wired interfaces.

\section{Acknowledgments}
This work was funded by the German Research Foundation (DFG) Priority Program SPP 2253 Nano Security.


\bibliographystyle{IEEEtran}
\bibliography{abbrev3,crypto,bibliography}

\end{document}